\documentclass[11pt]{article}
\usepackage{url} 
\usepackage[letterpaper, margin=0.95in]{geometry}
\usepackage{multicol}
\usepackage{dcolumn}
\newcolumntype{d}[1]{D{.}{.}{#1}}
\usepackage{siunitx}
\usepackage{graphicx}
\usepackage{orcidlink}
\usepackage{authblk}
\usepackage{amsmath,amssymb}
\usepackage{natbib}
\begin{document}

\title{Towards Three Component Seismograms From One Component DAS Records; Finite Frames in Geophysics}

\author[1,2]{Franklin G. Horowitz\, \orcidlink{0000-0001-8352-2602}}

\affil[1]{Horowitz Consulting}{}
\affil[2]{Institute for the Study of the Continents, Dept. EAS, Cornell University}{}

\maketitle

\begin{abstract}
  Some geophysical observations commonly collect only one component
  (1C) of a three component (3C) vector field. For example,
  Distributed Acoustic Sensing (DAS) records seismograms derived from
  displacement differences along the axis of segments of a fiber optic
  cable. In practice, multiple observations from such 3C vector fields
  are available, but commonly along non-orthogonal directions --
  i.e.\ desirable sets of observations along orthogonal basis vectors
  are not available. For DAS, the theory of (finite) frames allows the
  recovery of 3C vector observations as long as the set of
  measurements occur on axis-vectors that mathematically span 3D
  space.  A reconstruction algorithm from finite frame theory is
  described and then applied to geometry data from a borehole at the
  Utah FORGE geothermal project. The results demonstrate recovery of
  high precision 3C vectors along the fiber optic cable from 1C
  input vector-projections.
\end{abstract}

\section{Introduction}\label{Intro}
Recent advances in Distributed Acoustic Sensing (DAS) are enabling
many new applications in the dense collection of seismic data
\citep[e.g.][]{DASMonograph2021,Lindsey2021,LiMellorsZhan25}. However, a
major constraint of the prevalent (Rayleigh scattering) DAS technique
is that it only collects displacement information projected onto a set
of vectors along segments on the axis of the optical fiber
(axis-vectors). Such data are restricted to one component (1C) of a
three component (3C) particle displacement field. 

This present work aims to exploit the non-collinear nature of the
observable axis-vectors to recover estimates of the 3C particle
displacement vectors. In essence, this technique can estimate the 3
vector components because the axis-vectors can span 3D space in a
linear algebraic sense --- even though the axis vectors are not
orthonormal.

Previous approaches to solving this problem involve designing fiber
cable layout geometries --- such as spirals --- that explicitly
construct orthonormal components to be recovered at certain locations
\citep[e.g.][]{LiLiEtAl2024}.

At its heart, the technique in this paper is nothing more exotic than
recovering a vector from projections along a set of basis functions
(vectors, in our specific case). The elaborate mathematics simply
allows for more general abstractions (and their mathematical proof
``once and for all'') as well as for the over-complete set of
non-orthogonal basis vectors. But the essence is just vector
reconstruction from measured vector components --- similar to the
methods taught in secondary school.

This work requires one major approximation. DAS data are formally
``strains'' (respectively ``strain rates'' after time differentiation)
-- hence ostensibly one component of a six-independent-component
strain(rate) tensor field. There have been efforts to model the
mechanics of optical fibers
\citep[e.g.][]{ChapeleauBassil2021}. However, the observation physics
is fundamentally the action of a \emph{small amplitude} 3C vector
particle displacement (respectively velocity) seismic wave-field
acting on the endpoints of segments of the optical fiber inducing an
elongation or shortening \emph{strain} on that segment. Those are
measured as the path-length changes (i.e.\ optical phase angle changes
between ``reflections'' from scatterers) along the fiber.  To simplify
the tensor nature of the problem --- rather than modeling the full
mechanics of an optical fiber, in cladding, perhaps attached to a
rock-mass or well casing with variable coupling along the length of
the fiber --- I choose to approximate the problem as a particle
displacement field being passively sampled by endpoints of segments of
the optical fiber.  This approach should allow for variations of the
technique described in this work to be applied to other observations
of geophysical interest --- such as multiple look-direction
observations of the surface displacement field from InSAR.

The mathematical technique used here is called the ``Theory of Finite
Frames'' \citep[e.g.][for a readable overview]{CasazzaEtAl2013}. As
detailed in \citet{CasazzaEtAl2013} it is
originally derived from work of \citet{Gabor1946}, and put into a
modern framework by \citet{Daubechies1986}. The theory of frames is
quite general and underlies the mathematical theory of wavelets
\citep[e.g.][]{Daubechies1992, Kaiser}. For the purposes of this work, a
major attraction of the theory is that it can take advantage of an
over-complete set of vectors --- i.e.\ use more observations than 3 to
span 3D space.

\section{Method}\label{method}
This work is focused on vector geometry and not on the details of wave
propagation that are the traditional domain of seismology. However, to make a
formal connection with seismology, a very simple
wave is first described as an example. \citet[page
126]{AkiRichards2002} denote a ``separation of variables'' style
expression for a simple plane wave propagating in the \(\mathbf{k}\)
direction with the form:
\begin{equation} \label{PlaneWave}
  \exp[j(\mathbf{k}\cdot\mathbf{X} - \omega t)].
\end{equation}
Here, \(\mathbf{k} = (k_x,k_y,k_z)\) is the vector wave-number for the
wave, \(\mathbf{X}\) is the 3D Cartesian coordinate system, \(\omega\)
is an (angular) frequency, \(t\) is time, and \(j =
\sqrt{-1}\). 

This work concentrates on estimating/reconstructing 3C \(\mathbf{k}\)
components from the 1C recordings available from DAS.\ The utility of
this work is because \(\mathbf{k}\) is very unlikely to be exactly
parallel to the axis vectors of the DAS cable --- along which the 1C
information is recorded. Indeed, information about the full 3C of
\(\mathbf{k}\) is an example of the fundamental reason that standard
3C seismometers are deployed.

In the following, the direction of \(\mathbf{k}\) from
equation~\ref{PlaneWave} is a particular example of what will be
called \(\mathbf{x}\) --- the time-varying full 3C wave-field over 3D
space.

All the rest of seismology --- other than resolving the components
arriving along the \(\mathbf{k}\) vector --- is not considered in this
work. That remains the domain of standard seismology.

\begin{figure*}
  \centering
  \includegraphics[alt={Sketch of geometry under study. Short, linear,
    well survey segment endpoints, segments, and unit vectors ``phi
    sub i'' are shown with exaggerated angles between segments. Also
    shown are a simple example of an incoming seismic wavefield
    particle motion vector, the ``boxcar patch'' definition as a
    contiguous subset of an odd number of survey segments, and that
    the center segment of a boxcar is where the reconstructed 3C
    vector is
    located.},width=.95\textwidth]{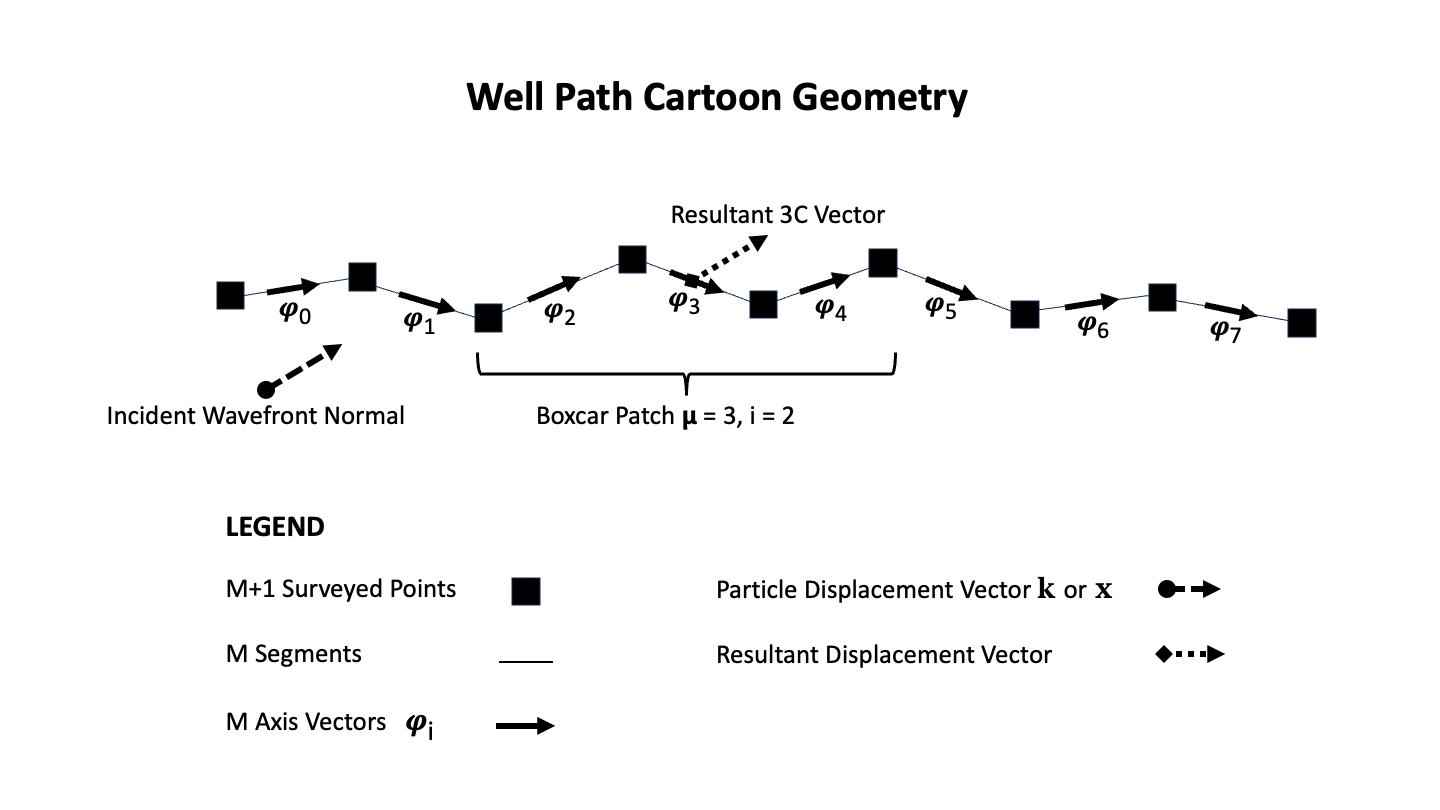}
  \caption{Cartoon geometry of the problem at hand. Refer to the main
    text for the symbol definitions. 
  }\label{Cartoon}
\end{figure*}

Figure~\ref{Cartoon} is a cartoon of the relevant geometry. Referring
back to this figure might be beneficial in reading the rest of this
paper.

\subsection{Finite Frames}\label{FiniteFrames}

The primary reference is \citet{CasazzaEtAl2013}, and (minor
variations of) their notation is adopted herein. They offer
mathematical proof of their results, so their results reported here
are well established. Selected portions of their work essential to
implementing the technique are quoted verbatim here. I do this in an
effort to make this paper self-contained for the geophysical community
interested mostly in applying their techniques --- as opposed to
gaining a deep understanding of the mathematical details, for which
interested readers should consult \citet{CasazzaEtAl2013} and the
references therein. The only claim to originality in this paper is in
applying those techniques to estimating 3C DAS signals from 1C DAS
observations. 

In this present paper, the general finite dimensional analysis of
\citet{CasazzaEtAl2013} is specialized to 3D physical space, and the
vector signals (i.e.\ the particle displacements due to arriving
seismic wave-fields) are assumed to be square-integrable. That is, the
(3D) spatially-extensive three-component displacement vector field --- denoted as
\(\mathbf{x}\) --- lies in 3 dimensional, real, Hilbert space
(\(\mathbf{x} \in \mathcal{H}^3 \)).

Applying their notation to the current work, the 3C (unit)
axis-vectors physically located at the center of the position of each
fiber segment, and locally aligned along the fiber (described above),
are indexed by \(i\) and are denoted \(\varphi_i\). See
Figure~\ref{Cartoon} for an illustration.  Also, to connect with the
previous simple plane wave example, the direction of \(\mathbf{k}\)
from equation~\ref{PlaneWave} is everywhere parallel to \(\mathbf{x}\)
in the current notation. Decomposition (projection) of \(\mathbf{x}\)
onto the \(\varphi_i\) is denoted by the map
\begin{equation} \label{Mapping}
  \mathbf{x} \mapsto \left(<\mathbf{x},\varphi_i> \right)_{i=1}^M
\end{equation}
where \(<,>\) denotes an inner (dot in our 3D case) product. To make
this abstract notion concrete, the projections on the RHS of
equation~\ref{Mapping} \emph{fully represent and denote} the multiple
seismic traces recorded along the DAS cable at any specified moment in
time at \(M\) distinct locations --- from the arrival of a seismic
wave-field such as the example of equation~\ref{PlaneWave}.

To establish that the \(\left(\varphi_i\right)_{i=1}^M\) is a frame
requires some technical details that are too far afield for this
paper. Those details and some of their consequences are described in
Definition 14 (equation 2, and following) of \citet{CasazzaEtAl2013} and
the interested reader should consult that work. Suffice it to say for
our purposes that the primary way for the
\(\left(\varphi_i \right)_{i=1}^M\) to fail the frame criteria is for
many of the \(\varphi_i\) to be exactly parallel or co-planar. Indeed,
a corollary of the frame property is that the
\(\left(\varphi_i \right)_{i=1}^M\) are a frame \emph{if and only if} they
are a spanning set of \(\mathcal{H}^3\)~\cite[Lemma 2 (ii),
p. 14]{CasazzaEtAl2013}. 

So far, the abstract claim is that the frame
\(\left(\varphi_i\right)_{i=1}^M\) is useful, and forms a spanning set
of \(R^3\) for (square-integrable) seismic waves. But actual
computation is required to do something concrete with this whole
approach. To do so, one first defines a so-called \emph{analysis}
operator \(T\):
\begin{equation} \label{AnalysisOp}
T\mathbf{x} := \left(<\mathbf{x},\varphi_i>\right)_{i=1}^M, \quad
\mathbf{x} \in \mathcal{H}^3
\end{equation}
\cite[Definition 15, p. 17]{CasazzaEtAl2013}.

In the current finite dimension work, the analysis operator \(T\)
defined in equation~\ref{AnalysisOp} can be represented as a matrix
\citep[e.g.][Definition 4, p. 7]{CasazzaEtAl2013}. \(T\) has an
associated \emph{synthesis} operator \(T^*\) --- given by the adjoint
(conjugate transpose matrix) of \(T\)~\cite[Definition 16,
p. 18]{CasazzaEtAl2013}. After these preliminaries, the connection is
made between the \(\varphi_i\) and the \(3 \times M\) matrix \(T^*\)
given by:
\begin{equation} \label{T*Matrix}
  T^* =   
  \begin{bmatrix}
    &| &| &\cdots &|    \\
    &\varphi_1 &\varphi_2 &\cdots &\varphi_M    \\
    &| &| &\cdots &|   \\
\end{bmatrix}
\end{equation}
\cite[Lemma 5, p. 19]{CasazzaEtAl2013}. Note that since \(\varphi_i
\in R^3\) no complex conjugation is necessary to form the adjoint
\(T^*\), just the transpose.

The frame operator \(S\) is defined by:
\begin{equation} \label{FrameOp}
  S\mathbf{x} := T^* T \mathbf{x} = \sum_{i=1}^M
  <\mathbf{x},\varphi_i>\varphi_i, \quad \mathbf{x} \in
  \mathcal{H}^3
\end{equation}
\cite[Definition 20, p. 20]{CasazzaEtAl2013}. Note that the second
equality here is a projection of the spatially extensive 3C signal
\(\mathbf{x}\) on to the \(\varphi_i\) to find coefficients, then
summing those coefficients multiplying their corresponding vectors
\(\varphi_i\) to form a resultant vector.  This construction
explicitly samples the full wave-field \( \mathbf{x} \) at the
locations of the basis vectors
\(\left(\varphi_i\right)_{i=1}^M\). Once again, see Figure~\ref{Cartoon} for the geometry. To implement the technique, both the
matrix form of the operator \(S\), and the explicit vector-component
projection and vector-summed resultant form of equation~\ref{FrameOp},
are required.

One finds the exact reconstruction formula by left-multiplying
equation~\ref{FrameOp} with \(S^{-1}\) while accounting for the fact
that \(S\) is a \(3 \times 3\) matrix in this work:
\begin{equation} \label{ExactReconstruction}
  x = \sum_{i=1}^M <\mathbf{x},\varphi_i> S^{-1} \varphi_i
\end{equation}
\cite[the first equality of Theorem 8, p. 25]{CasazzaEtAl2013}. Here,
I use the non-boldfaced vector \(x\) on the LHS of the equality to
denote a \emph{single} 3C resultant vector --- as opposed to the full
vector field \(\mathbf{x}\). To my reading, the theory stands mute on
the spatial location of \(x\) in this instance, but intuitively I
choose to assign the location to the center of a collection of
axis-vector segments being analyzed.

Now numerically speaking, the matrix (operator) \(S\) defined in
equation~\ref{FrameOp} might be ill-conditioned, because the
\(\varphi_i\) might well be ``nearly'' collinear or co-planar for any
given DAS deployment. So, inverting \(S\) to form \(S^{-1}\) --- as
required by equation\ \ref{ExactReconstruction} --- might be
numerically unstable. Reconstruction in such problematic cases might
be aided by using some iterative algorithms described in
\citet{CasazzaEtAl2013}.

\subsection{Practicalities}\label{Practicalities}

The novelty of this approach lies in the reconstruction of vector
components significantly different from those being observed along
the axis of a fiber-optic cable. As above, all of the usual problems of 3C
array seismology (e.g.\ move-out, migration, heterogeneous media, etc.)
are out-of-scope for this work. Demonstrating the recovery of 3
orthogonal components is the goal.

Clearly that goal, of estimating a single 3C vector from the layout of
an entire DAS array --- as suggested by equation~\ref{ExactReconstruction}
--- is less useful than estimating a collection of 3C vectors from the
interior of the array.

To that end --- and inspired by a moving average filter, sometimes
called a boxcar filter --- define a ``boxcar'' patch of length \(\mu\)
from adjacent subsets of the \(\varphi_i\), with \(\mu < M\). Assume
that the index \(i\) in the equations of section~\ref{FiniteFrames}
increments along consecutive axis-vectors \(\varphi_i\) of the
fiber-optic cable. For each index \(i < (M-\mu)\) define a different
boxcar patch. Define \(\mu\) to be an odd integer, and treat the
remaining number of axis-vectors on either side of the center segment
of the patch as the extra locations being used to estimate the 3C data
from the 1C observations. As long as \(\mu \ge (N = 3)\) for this 3C
problem, there are (at least) enough constraints to solve for the 3
unknown components. Now, all of the constructions of
section~\ref{FiniteFrames} carry through to each ``boxcar'' if \(M\)
is replaced by \(\mu\). To be explicit, the boxcar ``rolls along'' the
axis-vectors recovering a different 3C vector at each new
location. Once again, Figure~\ref{Cartoon} might provide some
geometric intuition for this procedure.

Assign the resulting 3C vector from equation~\ref{ExactReconstruction}
to the location of the center segment of each boxcar. If
reconstruction is successful, the result is an array of 3C vectors in
most of the interior of the fiber optic cable from the full
set of 1C observations.

Note that the exact reconstruction calculation of
equation~\ref{ExactReconstruction} needs only to be performed once to
be used throughout the time-span of the DAS cable deployment. To make
this explicit, here is the reconstruction expression as a function of
time:
\begin{equation} \label{preprocessing}
  x(t)|_{\ell_c} = \sum_{i=\ell}^{\ell + \mu}
  <\mathbf{x}(t),\varphi_i>\  S^{-1}_{\ell_c} \varphi_i .
\end{equation}
Here, \(x(t)|_{\ell_c}\) denotes 3C vector recovery through time at
the location \(\ell_c \equiv (\ell + int(\mu/2)+1)\) --- which is the
center index of a boxcar segment that starts at index \(\ell\) --- and
the function \(int\) is truncation to an integer.  Note that
\(S^{-1}_{\ell_c}\varphi_i\) is not a function of time (only of
position \(\ell_c\)). Strictly, the vector directions \(\varphi_i\)
might be considered as a function of time --- due to the rock-mass
strain. Assuming the directions to be independent of time is another
approximation consistent with the earlier approximation of using
displacement instead of strain described in 
section~\ref{Intro}.

Because the only functions of time on the RHS of
equation~\ref{preprocessing} are the DAS measurements
\(<\mathbf{x}(t),\varphi_i>\), each boxcar's
\(S^{-1}_{\ell_c}\varphi_i\) can be pre-computed and this entire
method can be regarded as a ``preprocessing'' step for recovering 3C
seismograms from DAS.

This boxcar approach is related to the concept of Fusion Frames in the
literature \citep{CasazzaEtAl2007, CasazzaEtAlFusionFrames2008,
  CasazzaEtAl2013, CasazzaKutyniok2013}, where results on robustness
to erasure are proven and used.

\section{Example Application}\label{Example} This technique
intrinsically depends on knowing the locations and orientations of the
fiber optic cable's axis-vectors. As an example, the publicly
available well-survey for the Utah FORGE \citep[e.g.][]{JonesEtAl2024}
borehole number 56-32 is used here \citep{Pankow2022}. The survey
consists of 97 down-hole points, which define 96 segments. The
axis-vectors \(\varphi_i\) are computed as unit vectors at the center
of each segment.  The frame technique from section~\ref{FiniteFrames}
-- slightly modified by the boxcar approach discussed in 
section~\ref{Practicalities}
-- is applied against
those data, as if a DAS survey were performed with a fiber-optic cable
following the well trajectory.

\begin{figure}
  \centering
  \includegraphics[alt={A) A plot of the well trajectory at the
    correct aspect ratio. The well is nearly vertical, suggesting that
    horizontal components would be difficult to recover. Also the
    horizontal test vector from section~\ref{Example} is
    displayed.  B) A plot of the well trajectory after horizontal
    exaggeration (vertical compression). Significant deviations --- on
    the order of 100 ft. --- from vertical are visible in the well
    segments.},width=.95\textwidth]{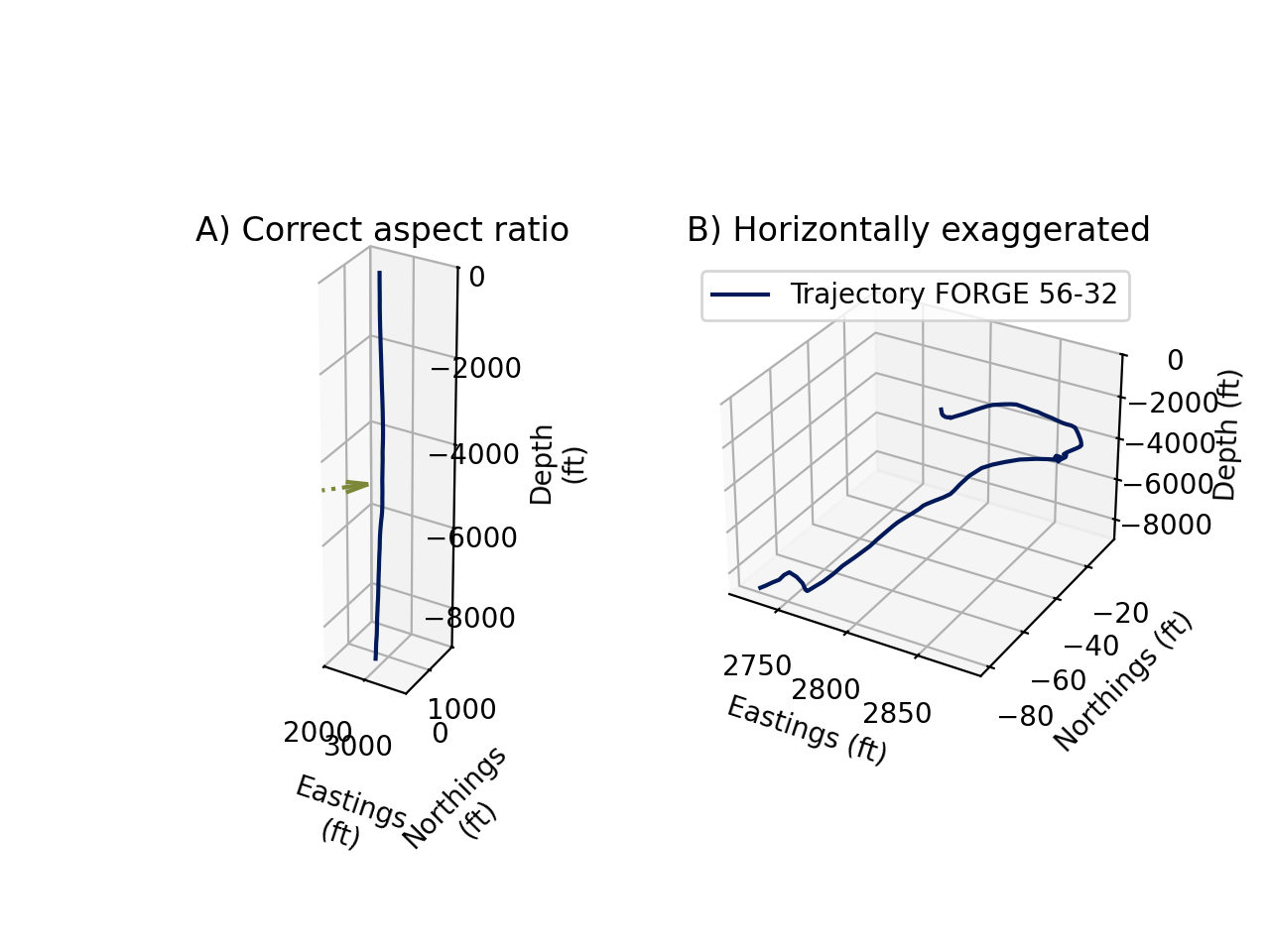}
  \caption{A) A plot of the well survey at the correct aspect ratio.
    B) A horizontally exaggerated (vertically compressed) plot of the
    same data. Note the small deviations in the survey. The well is
    essentially vertical.  Hence, the ability to recover horizontal
    vector components from measurements along its segments is much
    more problematic than recovering vertical components. The
    horizontal gray dotted arrow in A) shows the orientation of the test
    vector \((1\, 1\, 0)^T\) used in section~\ref{Example}.
    That test
    vector and orientation is used throughout the reconstruction shown
    in Table~\ref{VectorRecovery}.  }\label{NoVertExag}
\end{figure}

Figure~\ref{NoVertExag} shows the well trajectory from its
survey. Figure~\ref{NoVertExag}A is plotted at the correct aspect
ratio. To the eye, it looks essentially vertical --- and it is not
unreasonable to expect such geometry to be unable to resolve
horizontal components of incident vector
wave-fields. Figure~\ref{NoVertExag}B is plotted horizontally
exaggerated (vertically compressed) to emphasize the actual deviations
from vertical. It is all of those deviations that are being exploited
by the finite frame technique.

\begin{figure}
  \centering
  \includegraphics[alt={A histogram of inter-segment angles computed
    from the well trajectory of figure~\ref{NoVertExag} is
    shown. Angles are computed from the ArcCos of the dot-product of
    unit vectors along the segments. This surpresses the 3D nature of
    the inter-segment bends, which are crucial to the success of the
    method. Also, a CDF of the same data are shown. Fifty percent of
    the angles are 0.6 degrees or
    smaller.},width=.95\textwidth]{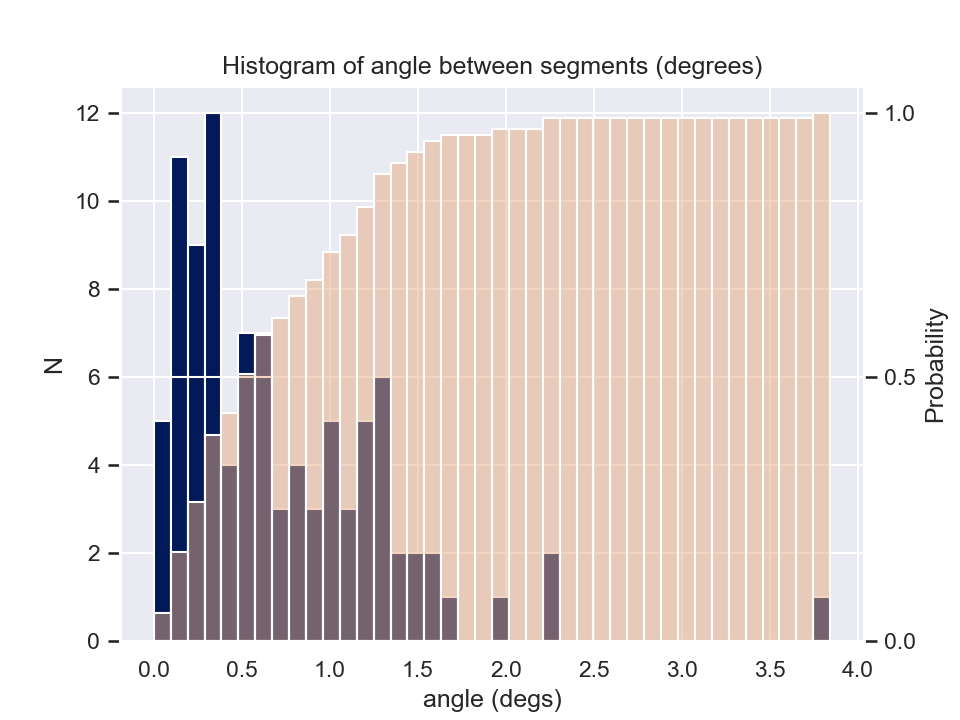}
  \caption{Angles (in degrees) between individual segments of the well
    survey shown in Figure~\ref{NoVertExag}. The angles are calculated
    from \(\cos^{-1}( <\varphi_{i+1},\varphi_i>)\) so orientation
    information in 3D space is not preserved. The standard histogram
    is plotted in dark blue against the left ordinate. Its CDF is
    plotted in a translucent tan against the right ordinate. A subdued
    red-tinted gray is where the two different kinds of histogram bins
    overlap.  Note that 50\% of the well survey segments have
    inter-segment angles of \(\le 0.6^\circ\).  The 3D nature of those
    angles is what is being exploited by the technique described in
    this paper.
  }\label{AngleHistogram}
\end{figure}

Figure~\ref{AngleHistogram} shows a histogram of the angles between
individual well-survey segments. From inspection, the maximum angle is
\(3.8^\circ\) and that occurs between only a single pair of
segments. Most of the inter-segment angles are much smaller. Note that
the Easting and Northing information is suppressed by this histogram.

As a proof of concept, the two horizontal elementary unit vectors in
the East and North directions are vector summed and projected against
the (essentially vertical) well survey segments described above. That
is, a constant-through-time test signal vector
\(\mathbf{x}(t) = (1\ 1\ 0)^T\) is inserted into
equation~\ref{preprocessing} at all locations \(\ell_c\). (See
Figure~\ref{NoVertExag}A for the geometry).  Reconstruction of all
three of those components demonstrates the technique.

For a boxcar length of \(\mu=7\), the computed results in
Table~\ref{VectorRecovery} demonstrates numerical recovery of those
two horizontal components to better than 6 significant digits, while
the vertical component is many orders of magnitude lower.

Note that the \(z\) components listed in Table~\ref{VectorRecovery}
are all from an incoming test signal vector --- which is explicitly
constructed to have zero in the \(z\) direction. Because the
calculations are performed with double-precision floating point
variables, the reported \(z\) components are due to round-off error
from cancellations in that component from summations of the
\(S^{-1}_{\ell_c} \varphi_i\) in equation~\ref{preprocessing}. I
attribute the reason that the \(x\) and \(y\) components are showing
such high (apparent) precision to some interplay between two effects:
Firstly, the Python code's pretty-printing algorithm does some
(opaque) internal calculation to display the result.  Secondly, the fact that
even in the double-precision floating point representation used in the
numerical calculation, the sum \(1 + \sim 10^{-16} \approx 1\) due to the
inability of floating point to represent such numbers exactly.

Table~\ref{VectorRecovery} demonstrates the accuracy and precision
available from this technique.

\begin{table*}
\centering
{\footnotesize
\renewcommand{\arraystretch}{0.39} 
\addtolength{\tabcolsep}{-2.5em}
\begin{minipage}{0.45\textwidth}
\begin{tabular}[t]{d{1} d{-1} d{-1} d{-1} }
\hline
 \mathrm{Index}& \mathrm{x}.\, \mathrm{component} & \mathrm{y}.\, \mathrm{component} & \mathrm{z}.\, \mathrm{component}  \\
\hline \\[0.1ex]
  3 & 1.00000e\!+\!00 & 1.00000e\!+\!00 &4.46122e\!-\!18\\
  4 &  1.00000e\!+\!00 &  1.00000e\!+\!00 &-1.73472e\!-\!18\\
  5 &  1.00000e\!+\!00 &  1.00000e\!+\!00 &-6.93889e\!-\!18 \\
  6 & 1.0000e\!+\!00 & 1.0000e\!+\!00 &1.1189e\!-\!16 \\
  7 &  1.00000e\!+\!00 &  1.00000e\!+\!00 &-1.04083e\!-\!16 \\
  8 & 1.00000e\!+\!00 & 1.00000e\!+\!00 & 3.10862e\!-\!15 \\
  9 &  1.00000e\!+\!00 &  1.00000e\!+\!00 &-1.79023e\!-\!15 \\
  10 & 1.00000e\!+\!00 & 1.00000e\!+\!00 &1.52656e\!-\!16 \\
  11 &  1.00000e\!+\!00 &  1.00000e\!+\!00 &-1.38778e\!-\!17 \\
  12 & 1.00000e\!+\!00 & 1.00000e\!+\!00 &4.85723e\!-\!16 \\
  13 &  1.00000e\!+\!00 &  1.00000e\!+\!00 &-3.81639e\!-\!16 \\
  14 & 1.00000e\!+\!00 & 1.00000e\!+\!00 &3.60822e\!-\!16 \\
  15 & 1.0000e\!+\!00 & 1.0000e\!+\!00 &1.9082e\!-\!16 \\
  16 & 1.00000e\!+\!00 & 1.00000e\!+\!00 &4.16334e\!-\!16 \\
  17 &  1.00000e\!+\!00 &  1.00000e\!+\!00 &-1.97758e\!-\!16 \\
  18 & 1.0000e\!+\!00 & 1.0000e\!+\!00 &3.1225e\!-\!17 \\
  19 &  1.00000e\!+\!00 &  1.00000e\!+\!00 &-5.81132e\!-\!17 \\
  20 & 1.00000e\!+\!00 & 1.00000e\!+\!00 &2.80331e\!-\!15 \\
  21 & 1.00000e\!+\!00 & 1.00000e\!+\!00 &8.88178e\!-\!16 \\
  22 & 1.00000e\!+\!00 & 1.00000e\!+\!00 &7.49401e\!-\!16 \\
  23 &  1.00000e\!+\!00 &  1.00000e\!+\!00 &-5.82867e\!-\!16 \\
  24 &  1.0000e\!+\!00 &  1.0000e\!+\!00 &-5.6205e\!-\!16 \\
  25 & 1.00000e\!+\!00 & 1.00000e\!+\!00 &8.25728e\!-\!16 \\
  26 & 1.0000e\!+\!00 & 1.0000e\!+\!00 &2.4503e\!-\!16 \\
  27 &  1.00000e\!+\!00 &  1.00000e\!+\!00 &-6.93889e\!-\!16 \\
  28 & 1.00000e\!+\!00 & 1.00000e\!+\!00 &1.38778e\!-\!17 \\
  29 & 1.00000e\!+\!00 & 1.00000e\!+\!00 &6.59195e\!-\!17 \\
  30 &  1.00000e\!+\!00 &  1.00000e\!+\!00 &-2.60209e\!-\!18 \\
  31 & 1.00000e\!+\!00 & 1.00000e\!+\!00 &5.89806e\!-\!17 \\
  32 & 1.00000e\!+\!00 & 1.00000e\!+\!00 &1.17961e\!-\!16 \\
  33 & 1.00000e\!+\!00 & 1.00000e\!+\!00 &3.20924e\!-\!17 \\
  34 & 1.00000e\!+\!00 & 1.00000e\!+\!00 &3.81639e\!-\!17 \\
  35 &  1.00000e\!+\!00 &  1.00000e\!+\!00 &-7.80626e\!-\!18 \\
  36 & 1.00000e\!+\!00 & 1.00000e\!+\!00 &1.51788e\!-\!18 \\
  37 &  1.00000e\!+\!00 &  1.00000e\!+\!00 &-1.38778e\!-\!17 \\
  38 &  1.00000e\!+\!00 &  1.00000e\!+\!00 &-8.67362e\!-\!18 \\
  39 & 1.00000e\!+\!00 & 1.00000e\!+\!00 &1.73472e\!-\!18 \\
  40 &  1.00000e\!+\!00 &  1.00000e\!+\!00 &-1.73472e\!-\!18 \\
  41 &  1.00000e\!+\!00 &  1.00000e\!+\!00 &-3.03577e\!-\!18 \\
  42 & 1.00000e\!+\!00 & 1.00000e\!+\!00 &1.73472e\!-\!18 \\
  43 &  1.00000e\!+\!00 &  1.00000e\!+\!00 &-2.60209e\!-\!17 \\
  44 &  1.00000e\!+\!00 &  1.00000e\!+\!00 &-4.22839e\!-\!18 \\
  45 &  1.00000e\!+\!00 &  1.00000e\!+\!00 &-8.67362e\!-\!19 \\
  46 & 1.00000e\!+\!00 & 1.00000e\!+\!00 &5.63785e\!-\!18 \\
  47 & 1.00000e\!+\!00 & 1.00000e\!+\!00 &3.46945e\!-\!18 \\
  \vdots & \vdots & \vdots & \vdots\\
\\[0.1ex]\hline
\end{tabular}
\end{minipage}  \hspace{1em}
\begin{minipage}{0.45\textwidth}
\begin{tabular}[t]{d{3} d{-1} d{-1} d{-1} }
\hline
 \mathrm{Index}& \mathrm{x}.\, \mathrm{component} & \mathrm{y}.\, \mathrm{component} & \mathrm{z}.\, \mathrm{component}  \\
  \hline\\[0.1ex]
  \vdots & \vdots & \vdots & \vdots\\
  48 & 1.00000e\!+\!00 & 1.00000e\!+\!00 &9.54098e\!-\!18 \\
  49 & 1.00000e\!+\!00 & 1.00000e\!+\!00 &6.50521e\!-\!18 \\
  50 & 1.00000e\!+\!00 & 1.00000e\!+\!00 &3.46945e\!-\!18 \\
  51 & 1.00000e\!+\!00 & 1.00000e\!+\!00 &3.68629e\!-\!18 \\
  52 &  1.00000e\!+\!00 &  1.00000e\!+\!00 &-2.38524e\!-\!18 \\
  53 &  1.00000e\!+\!00 &  1.00000e\!+\!00 &-8.67362e\!-\!19 \\
  54 & 1.00000e\!+\!00 & 1.00000e\!+\!00 &1.82146e\!-\!17 \\
  55 & 1.00000e\!+\!00 & 1.00000e\!+\!00 &1.04083e\!-\!17 \\
  56 &  1.0000e\!+\!00 &  1.0000e\!+\!00 &-2.1684e\!-\!17 \\
  57 &  1.00000e\!+\!00 &  1.00000e\!+\!00 &-4.16334e\!-\!17 \\
  58 &  1.00000e\!+\!00 &  1.00000e\!+\!00 &-2.77556e\!-\!17 \\
  59 & 1.00000e\!+\!00 & 1.00000e\!+\!00 &1.52656e\!-\!16 \\
  60 &  1.00000e\!+\!00 &  1.00000e\!+\!00 &-4.85723e\!-\!17 \\
  61 &  1.00000e\!+\!00 &  1.00000e\!+\!00 &-7.35523e\!-\!16 \\
  62 & 1.00000e\!+\!00 & 1.00000e\!+\!00 &3.05311e\!-\!16 \\
  63 &  1.00000e\!+\!00 &  1.00000e\!+\!00 &-6.93889e\!-\!18 \\
  64 & 1.00000e\!+\!00 & 1.00000e\!+\!00 &6.59195e\!-\!17 \\
  65 &  1.00000e\!+\!00 &  1.00000e\!+\!00 &-6.41848e\!-\!17 \\
  66 & 1.00000e\!+\!00 & 1.00000e\!+\!00 &2.25514e\!-\!17 \\
  67 & 1.00000e\!+\!00 & 1.00000e\!+\!00 &5.55112e\!-\!17 \\
  68 & 1.00000e\!+\!00 & 1.00000e\!+\!00 &7.28584e\!-\!17 \\
  69 &  1.00000e\!+\!00 &  1.00000e\!+\!00 &-6.93889e\!-\!17 \\
  70 &  1.00000e\!+\!00 &  1.00000e\!+\!00 &-1.16226e\!-\!16 \\
  71 &  1.00000e\!+\!00 &  1.00000e\!+\!00 &-3.29597e\!-\!17 \\
  72 & 1.000e\!+\!00 & 1.000e\!+\!00 &6.245e\!-\!17 \\
  73 & 1.00000e\!+\!00 & 1.00000e\!+\!00 &3.46945e\!-\!17 \\
  74 & 1.00000e\!+\!00 & 1.00000e\!+\!00 &6.39679e\!-\!17 \\
  75 & 1.00000e\!+\!00 & 1.00000e\!+\!00 &1.00614e\!-\!16 \\
  76 & 1.00000e\!+\!00 & 1.00000e\!+\!00 &6.93889e\!-\!17 \\
  77 & 1.00000e\!+\!00 & 1.00000e\!+\!00 &1.38778e\!-\!17 \\
  78 & 1. & 1. &0. \\
  79 & 1.00000e\!+\!00 & 1.00000e\!+\!00 &-2.42861e\!-\!17 \\
  80 & 1.00000e\!+\!00 & 1.00000e\!+\!00 &9.36751e\!-\!17 \\
  81 & 1.00000e\!+\!00 & 1.00000e\!+\!00 &7.28584e\!-\!17 \\
  82 & 1.0000e\!+\!00 & 1.0000e\!+\!00 &3.1225e\!-\!17 \\
  83 &  1.00000e\!+\!00 &  1.00000e\!+\!00 &-7.02563e\!-\!17 \\
  84 & 1.00000e\!+\!00 & 1.00000e\!+\!00 &3.79471e\!-\!17 \\
  85 & 1.00000e\!+\!00 & 1.00000e\!+\!00 &8.32667e\!-\!17 \\
  86 & 1.00000e\!+\!00 & 1.00000e\!+\!00 &3.46945e\!-\!18 \\
  87 &  1.00000e\!+\!00 &  1.00000e\!+\!00 &-6.93889e\!-\!18 \\
  88 & 1.00000e\!+\!00 & 1.00000e\!+\!00 &5.20417e\!-\!18 \\
  89 &  1.00000e\!+\!00 &  1.00000e\!+\!00 &-1.56125e\!-\!17 \\
  90 & 1.00000e\!+\!00 & 1.00000e\!+\!00 &6.78711e\!-\!17 \\
  91 & 1.00000e\!+\!00 & 1.00000e\!+\!00 &4.85723e\!-\!17 \\
  92 & 1.00000e\!+\!00 & 1.00000e\!+\!00 &-9.71445e\!-\!17 \\
\\[0.1ex]\hline
\end{tabular}
\end{minipage}
}
\caption{Resultant vectors from the boxcar variant
  (equation~\ref{preprocessing}) for the survey from Utah FORGE Well
  56-32, with \(M=96\) and \(\mu=7\). The indices range over all valid
  locations (\(\ell_c\)) for these specified \(\mu\) and \(M\). The
  test vector \(\mathbf{x}\!=\!(1\,1\,0)^T\) at all locations. Recall
  that the well trajectory is essentially sub-parallel to \(z\) so the
  recovery of the \(x\) and \(y\) components and the suppression of
  the \(z\) component is remarkable.\label{VectorRecovery}}

\end{table*}

\section{Discussion}\label{Discussion}

The technique described here only deals with the geometry and math of
turning a set of 1C observations into 3C resultant vectors. The
temporal nature of the incident wave-field is explicitly out of scope
for this work. However, that temporal nature is the domain of standard
seismology, and all of the knowledge, techniques and experience of
seismologists clearly needs to be brought to bear on that part of the
problem. Think of this technique as merely constructing virtual 3C
(strain/strain rate) seismometers from 1C data, rather than using the
resulting 3C seismograms to analyze the geophysical problem of
interest.

This technique also has drawbacks. The primary drawback is the need to
spatially spread out the measurement procedure. This results in
non-localized measurements being concentrated into the resultant
vector \(x(t)|_{\ell_c}\) in equation~\ref{preprocessing}. The boxcar
approach spreads the 1C DAS measurements over \(\mu\) adjacent fiber-optic segments in
order to produce one 3C resultant vector. This attribute has a similar
character to that of (mathematical, not seismological) wavelets ---
which are already widely deployed in geophysics. So, dealing with such
problems might be familiar to the community.

In this seismological application, the spatial spreading out will
interact with the different arrival times from wave-fronts traveling at
angles to the fiber-optic cable segment. Depending on the DAS
interrogator's parameters, that may well cause difficulties with
temporal smearing of phase arrivals. Like a wavelet, one possible
approach to ameliorate that effect might be to concentrate the spatial
spreading (shorten the ``support'') by choosing a smaller value for
\(\mu\) in the boxcar approach.

Be aware, however, that using smaller \(\mu\) increases the chance of
using non-spanning sets of axis-vectors in equation~\ref{T*Matrix} ---
leading to \(S_{\ell_c}\) being numerically non-invertible in certain
locations. Indeed, when \(\mu=3\) or \(\mu=5\) this is observed at a
few locations with the well survey data from 
section~\ref{Example}.
The symptom of that problem is when a few locations
show drastically different vector components from the correct ones. In
a real world situation where the ``correct'' answer is unknown,
perhaps the best approach might be to increase \(\mu\) until the
vector components stop drastically changing. If the misbehaving
location happens to be at a crucial location, perhaps attempting one
of the iterative algorithms described by \citet{CasazzaEtAl2013} might
be worth considering.

Another practical aspect that potentially could be problematic. The
DAS gauge length --- a parameter which can be affected by settings on
the DAS interrogator --- is unlikely to be the same as the surveyed
segment size (e.g.) as shown in Figure~\ref{Cartoon}. This means that
the input DAS components used in \(<\mathbf{x},\varphi_i>\) in
equations~\ref{AnalysisOp},~\ref{ExactReconstruction} and~\ref{preprocessing} might need to be something like an average over
all gauge lengths in the given survey segment. In other words, it is
likely that signal processing creativity still will be required in
using this method.

Clearly, the technique is not limited to well-bore deployments of DAS. Any
(existing?) fiber optic cable deployment with good ground coupling and a
detailed positional survey would work. Indeed, this technique might
conceivably spur re-surveys of existing cable deployments simply to
take advantage of recovering 3C data. Existing (communications?)
fiber-optic systems in seismogenic places such as California, Japan, T\"urkiye,
New Zealand or Iceland would be prime candidates for such
retrospective surveys --- as an aid to monitoring seismicity.

I leave an analysis of directional error statistics
due to this technique for future work.

\section{Conclusion}\label{Conclusion}

Finite frames, an existing technique from the applied mathematical
literature, has been adapted to recovering 3C vectors from 1C vector
projections along a fiber optic cable with a detailed geometric
survey. In essence, the technique creates virtual 3C seismometers from
DAS data. Many seismological applications are possible --- as well as
other geophysical applications where only one vector component is
observed but recovery of the full three component vector is desired.

\section{Data and Software Availability}
The well data from \citet{Pankow2022} may be found at the Geothermal
Data Repository:
\url{https://gdr.openei.org/submissions/1440}.

The Python source code implementing the algorithm and creating the
figures and table data for this paper may be found in the git
repository
\url{https://fghorow@bitbucket.org/geothermalcode/jupyerlabdasframes.git}. It
is freely licensed under the Creative Commons License CC BY
4.0. Details of this license may be found at
\url{https://creativecommons.org/licenses/by/4.0/}.

\section{Declaration of Competing Interests}
The author declares there are no conflicts of interest for this manuscript.

\section{Acknowledgements}

I thank Tom Pratt of the USGS for a recent conversation that reminded
me of the applicability of this technique to DAS observations. I thank
Anastasia Stroujkova of Enegis for pointing out that my assumption of
the axis-vector directions being constant with time is only an
approximation. I thank Korin Carpenter for spurring me to investigate
the exact recovery algorithm when I had been assuming that one of the
iterative techniques would be required due to the ill-conditioned
nature of the \(S\) operator. I thank Pam Smith for editorial help
with early drafts of this document. Last but not least, I thank Les
Schaffer for an extensive discussion which resulted in significant
clarity improvements to this paper. Any errors remaining are mine
alone. I dedicate this work to the memory of my only child, Ian
B. Horowitz. This work was self-funded.

\bibliographystyle{agu}
\bibliography{./DASFramesArXiv.bbl}
\end{document}